# Geospatial Road Cycling Race Results Data Set

Bram Janssens[1,2,3], Luca Pappalardo[4], Jelle De Bock[5], Matthias Bogaert[1,2], Steven Verstockt[5]

**Affiliations**
1. Ghent University, Department of Marketing, Innovation, and Organization, Tweekerkenstraat 2, 9000 Ghent, Belgium
2. FlandersMake@UGent–corelab CVAMO
3. Research Foundation Flanders
4. ISTI-CNR, Via Moruzzi, 1, Pisa, 56127, Italy
5. University of Ghent - imec, IDLab; Technologiepark-Zwijnaarde 122, 9052 Gent, Belgium

Corresponding author: Bram Janssens (bram.janssens@ugent.be)

## Abstract

The field of cycling analytics has only recently started to develop due to limited access to open data sources. Accordingly, research and data sources are very divergent, with large differences in information used across studies. To improve this, and facilitate further research in the field, we propose the publication of a data set which links thousands of professional race results from the period 2017-2023 to detailed geographic information about the courses, an essential aspect in road cycling analytics. Initial use cases are proposed, showcasing the usefulness in linking these two data sources.

## 1. Introduction

Data volumes in professional road cycling are growing: more and more athletes use these devices to measure their performance before entering a race and make tactical decisions during a race. This data-driven type of racing is so omnipresent that it led to critique from fans and outsiders to the point that GCN, a popular news medium, created some content on the potential ban of power meters in the sport[1]. Nonetheless, these data-tracking devices are here to stay given the abundance of potential information they generate and the potential utility for academics and professionals. A first study by Cintia, Pappalardo and Pedreschi[2] uses data from the popular sports social media platform Strava to give some large-scale evidence to the effectiveness of polarized training in cycling with a sample of tens of thousands of amateur athletes.

However, since then, data-driven innovation in cycling analytics has been lagging behind on its potential. A major explanation lies in the growing sensitivity associated with personal data protection and privacy[3]. This has resulted in several incentives, with the European Union's General Data Protection Regulation (GDPR) perhaps the most known one. While the advantages of these initiatives are clear, they are also slowing down the sharing of vital data sources in research[4]. Accordingly, other solutions are searched within academia. A popular technique is data anonymization[5], where the information which could identify data objects (e.g., riders' identifiers or names) is removed or drastically reduced. This technique has been applied to open-source projects such as the GoldenCheetah Open Data project[6], which provides training information from users of the GoldenCheetah training platform. Containing anonymized data from thousands of amateur athletes using the platform, it has proven a valuable source of information to perform physiological studies on a much broader scale[7].

A similar database containing spatiotemporal training and race data from Strava records might be an interesting option to study professional athletes. However, such an anonymized data set cannot be created easily as the information on the ending position could easily lead to linkage attacks[5], which would lead to the unauthorized sharing of personal information on these athletes. Moreover, researchers cannot use truly personal data without explicit permission of athletes, leading to limited data samples of a few athletes[8] or researchers using fully publicly available sources. A considerable downside, however, is that these data sources are scattered around the internet, leading to studies which use different data sources[9–11].

Two attributes stand out in these studies: the dispersion of techniques and data sources and the limited inclusion of race course information. Road cycling is a unique sport, as the race course ridden heavily impacts the race favorites. A sprinter like Mark Cavendish is either expected to finish among the first when the race course is relatively flat, or to be battling against the time limit when the race enters the mountains. Accordingly, one should account for the race course.

So far, no study has used this information concerning performance evaluation (e.g., outcome estimation or talent identification). This can be explained by how each data source uses different names for the same races. These differences are never huge but they make it more challenging to combine these sources on a large scale. For instance, stages in the Critérium du Dauphiné will receive a *dauphine* tag on the ProCyclingStats[12] website, while the race is called *Criterium du Dauphine* on the La Flamme Rouge[13] platform. These differences are minor but further complicated by women editions of the same race or Junior or U23 versions. Moreover, races often change names throughout the years, with a famous example being the current Renewi Tour, which has also been called the Eneco Tour, the BinckBank Tour, and the Benelux Tour during recent years. The difficulties in merging course data with results data might explain why these combined data sources have been limitedly used in prior research.

To alleviate this issue and provide a standard ground on which future cycling studies can be developed, we propose a data set including detailed information on both race courses ridden, as well as race results linked to these courses. Our primary goal in releasing these data is to have a standardized data set which will lead to easier-to-compare studies, and lower the hurdle to start doing research in the field. The data set allows future researchers to develop new methodologies incorporating the well-discussed specializations within cycling[14]. Moreover, we show that the data bears the potential to boost the performance of several existing cycling analytics applications, as well as create new interesting applications. This could further boost data-driven innovation in the sport which, despite WorldTour budgets summing up to hundreds of millions of euros[15], lacks behind compared to other major European sports.

## 2. Methods

The La Flamme Rouge website provides community-created GPX files on a wide list of top-level road cycling races. The community copies the public information on road courses into GPX files. We build a web scraper to download all GPX files available on the website of all cycling seasons from 2017 to 2023, obtaining an overall sample of 8,092 races. Race organizers always share the route they will follow. The reason to opt for the secondary data source of La Flamme Rouge instead, is based on the fact that organizers typically share their course information in hard-to-process formats such as images or pdfs while being scattered across a wide range of sources. Rather, the GPX files from La Flamme Rouge are relatively easy to process, located at one single location, and are easily linked to additional geospatial information, as we do for our structured course information in this data set as described below.

Each individual GPX file is linked to the open source OpenStreetMap[16] database. This enables us to get detailed information about the surface and way types ridden, which gives detailed information about several road cycling segments that can decide races (e.g., cobblestones and gravel sections), but also about how wide the roads are, which in turn gives information about how important positioning skills are. The distance per road type and surface type is calculated, as well as other statistics such as elevation gain or total distance. The total list of crafted features is depicted in Table 2 of the related section. The data was automatically checked for infeasible values (e.g., race distances > 500 km, elevation gain in meters / distance in km > 100). These infeasible values were then manually checked and set to NA if necessary. The current version of the data still contains these missing values to enable the data set users to be as flexible as necessary. However, we already give some guidance to potential imputation techniques in Section 4 below.

The most value from the suggested data set is in its linkage with actual race results. De Bock and Verstockt[17] use fuzzy matching to match GPX files with results from the Union Cycliste Internationale (UCI) database based upon race name, stage number, and date. This approach is valuable but needs a clearly defined database upfront to do the matching with, which is not feasible for our approach. Rather, we search race results across various online sources using the Google search engine[18] and make a script to search for the related race using the name in the La Flamme Rouge database. As an additional check, dates are checked to fall within the same year, as well as the race result's reported distance to fall within a 10% margin of the GPX file distance. Typical issues are related to a misidentification of a youth result as an adult result, vice versa, or related to male/female misidentifications. Distance differences between these categories are typically larger than 10% (e.g., 2023 men's/women's editions of Tour of Flanders: 273 km vs. 157 km). Figure 1 displays how this information can easily be retrieved from websites displaying race results. The entire process was monitored throughout to ensure no false linkages. Overall, the procedure above resulted in 6,394 race results which are linked to our race courses. This number is smaller than the number of retrieved GPX files because not all race course files can be matched to results in the PCS database. Typically, the most important races (.WT and .Pro level) are included.

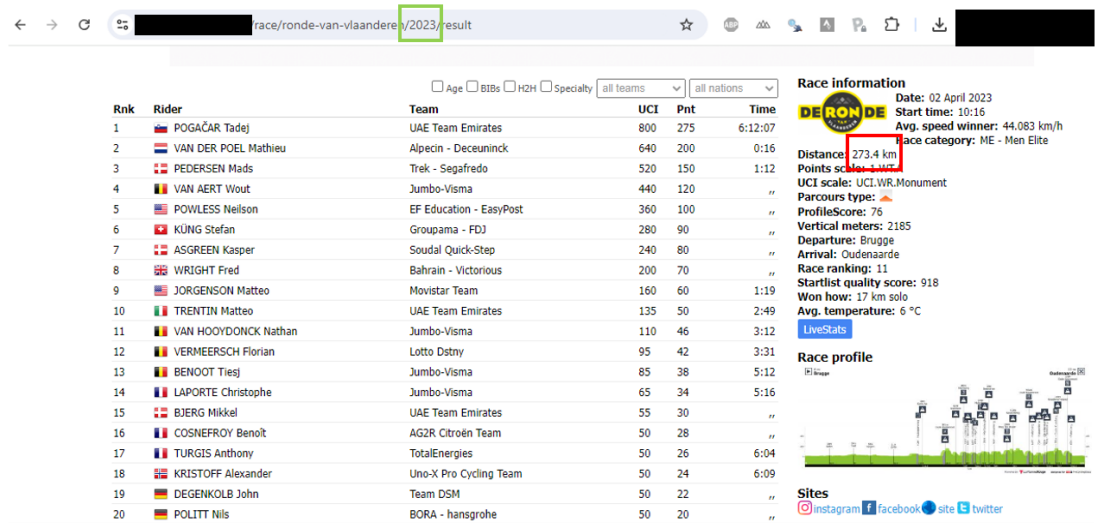

**Figure 1**: Example output (Tour of Flanders) from race results website. Green square indicates how URL structuring can be used to check for the correct year, and red bracket indicates easy detection of race distance in HTML.

This project got approved by the Ethics Committee of Ghent University's Faculty of Economics and Business Administration. The only data used about human subjects are their end results in professional competitions. These end results are shared by the race organizers and participating teams, after which several outlets, including both popular press as well as fan-based initiatives continue to disseminate this information to a wider audience. Beside this public information on the public role these athletes play, no information is collected about them individually.

## 3. Data Records

The data set consists of three parts, which will be outlined below. All data is available at Cycling Analytics Data Sets (figshare.com).

***GPX files.*** The GPX files contain information about the sequence of locations visited, with information on latitude, longitude, and elevation. An example of a part of a GPX trace is provided in Figure 2 and Table 1, before and after re-structuring, respectively. Many popular packages are available to re-structure GPX files in the format of Table 1, such as *gpxpy[19]* and *scikit-mobility[20]*. While containing only three columns (i.e., end users should disregard the incorrect time value), these features can quickly be converted into very useful information when using some differential calculations on them, or linking them to external geospatial data sources. For example, Figure 3 shows how you can match these data points to additional geospatial information.

```
        </trkpt>
        <trkpt lat="50.88767" lon="5.72682">
          <ele>47</ele>
          <time>2022-06-07T17:17:50+02:00</time>
        </trkpt>
        <trkpt lat="50.88792" lon="5.72663">
          <ele>47</ele>
          <time>2022-06-07T17:17:52+02:00</time>
        </trkpt>
        <trkpt lat="50.88917" lon="5.72599">
          <ele>46</ele>
          <time>2022-06-07T17:18:05+02:00</time>
        </trkpt>
        <trkpt lat="50.89017" lon="5.72549">
          <ele>45</ele>
          <time>2022-06-07T17:18:14+02:00</time>
        </trkpt>
        <trkpt lat="50.89042" lon="5.72538">
          <ele>45</ele>
          <time>2022-06-07T17:18:16+02:00</time>
```

**Figure 2**: GPX point sequences as included in GPX files (i.e., before re-structuring). Same example as used in Table 1 (i.e., 2017 Amstel Gold Race). Note how <time> contains no useful or correct information (race was actually in April 2017), and should be disregarded by end users. Time values are auto-generated by the platform (based on moment of collection) and incorrect.

**Table 1**: Example *gpx_files* data after re-structuring: Sequence of points visited during 2017 Amstel Gold Race. Note how the 'Time' column contains no useful or correct information (race was actually in April 2017), and should be disregarded by end users.

| Latitude | Longitude | Elevation |
|----------|-----------|-----------|
| … | … | |
| 50.88767 | 5.72682 | 47.0 |
| 50.88792 | 5.72663 | 47.0 |
| 50.88917 | 5.72599 | 46.0 |
| 50.89017 | 5.72549 | 45.0 |
| 50.89042 | 5.72538 | 45.0 |

| ... | ... | ... |
|---|---|---|

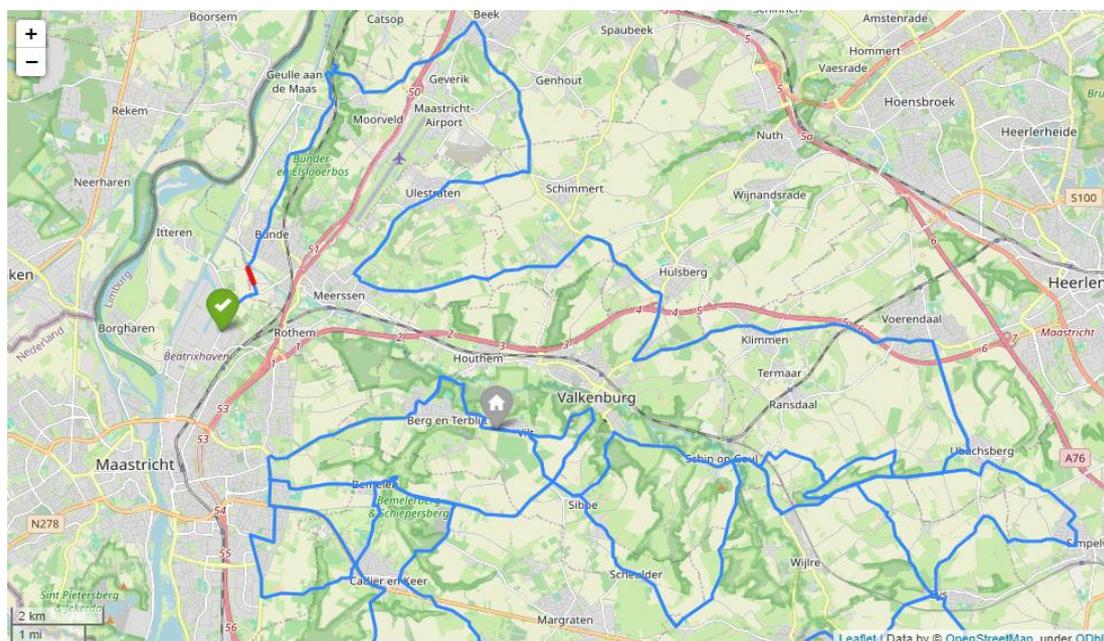

**Figure 3**: Visualization short sequence from Table 1 and Figure 2 (red segment) onto the entire route of 2017 Amstel Gold Race (blue trace).

***Course data.*** We also provide a comma-separated file that contains structured features about the race course, retrieved from the raw GPX files and OpenStreetMap. Table 2 gives an overview of the created features, while Table 3 displays several rows (with a subset of features) from the file. These features were used for our analyses below and can aid new researchers in getting started, but the raw GPX files provided allow for much more flexible analyses.

**Table 2**: Structured Course Features

| Feature | Description |
|---|---|
| **Race Name** | Name of the race. Can be matched exactly with GPX files by appending '.gpx' |
| **Highest Elevation** | Highest elevation point (meters) during course |
| **Lowest Elevation** | Lowest elevation point (meters) during course |
| **Net Gain** | Difference altitude end point (meters) compared to starting point |
| **Access Road** | Kilometers of access road: wide road type |
| **Alpine** | Kilometers of alpine hiking trails. Very uncommon. |
| **Asphalt** | Kilometers of asphalt |
| **Cobblestones** | Kilometers of cobblestones |
| **Compacted Gravel** | Kilometers of compacted gravel: intermediate between paved and loose gravel |
| **Cycleway** | Kilometers of cycle way: cycle tracks and lanes |
| **Off-grid (unknown)** | Kilometers of unregistered road |
| **Path** | Kilometers of path: very narrow road type |
| **Paved** | Kilometers unpaved underground |
| **Road** | Kilometers of road: intermediate width |
| **Singletrack** | Kilometers of single track: very narrow and unpaved |
| **State Road** | Kilometers of state road: wide road type |
| **Street** | Kilometers of street: narrow road type |
| **Unknown** | Kilometers of unknown underground |
| **Unpaved** | Kilometers unpaved underground |
| **Vertical Gain** | Total upwards elevation covered (meters) during course |
| **Downhill** | Total downwards elevation covered (meters) during course |
| **Distance** | Total distance covered (meters) |

The preliminary list of features in this file already gives an enormous depth of information to field experts, and potentially to intelligent models and algorithms. Consider Critérium du Dauphiné Stage 4 in Table 3: the lower distance clearly suggests a time trial, which can inform about very different race tactics, which will center purely around raw power, and self-pacing. The other races depicted from the Dauphiné have a more similar profile, especially when compared to the Omloop van het Hageland. This Belgian race is much flatter and one could expect this race to have a bunch sprint based on the elevation profile. However, this is where the other features give useful information: 82.8 km is raced on paved roads, and 68.7 km on cycle ways. This indicates a much more narrow and more twisty road structure which lowers the chances of a bunch sprint. Indeed, when looking at the results of this race, only 20 riders finished in the time of the first group (depicted in Table 5). Similarly, stage 2 of the Dauphiné stands out based upon its net gain, indicating that the finish was placed at much higher elevation when compared to the starting place. This also translates in the race having a much punchier winner (i.e., Julian Alaphilippe) compared to the other stages.

**Table 3**: Example *structured_course_data* data. Not all features (Table 2) are included to give an easier overview of the features which are relevant to the discussed differences across the selected races.

| Race Name | Distance | Road | Paved | Asphalt | Cycle-way | State Road | Net Gain | Vertical Gain | Downhill |
|---|---|---|---|---|---|---|---|---|---|
| 2023 Craywinckelhof - Omloop van het Hageland | 123.8 | 47.5 | 82.8 | 38.5 | 68.7 | 2.93 | 10 | 890 | 880 |
| 2023 Criterium du Dauphine Stage 1 | 158.73 | 158 | NA | 158 | NA | NA | -110 | 2890 | 3000 |
| 2023 Criterium du Dauphine Stage 2 | 167.9 | 140 | NA | 168 | NA | 27.9 | 680 | 2730 | 2050 |
| 2023 Criterium du Dauphine Stage 3 | 194.2 | 157 | 4.74 | 190 | 8.14 | 28.4 | -170 | 2210 | 2380 |
| 2023 Criterium du Dauphine Stage 4 | 31 | 31 | NA | 31 | NA | NA | -40 | 390 | 430 |
| 2023 Criterium du Dauphine Stage 5 | 190.59 | 160 | 1.45 | 189 | 1.2 | 29.2 | 160 | 1980 | 1820 |

***Race Results 2017-2023.*** We also link the data on the course profile with the related race results. Table 4 depicts the features present in this file, while Table 5 depicts the top results from the Omloop van het Hageland race discussed above. Note that both the official UCI and non-official PCS points are calculated due to the popularity of PCS points in previous research (e.g., [11]). Information about the rider's positioning, time differences, team, and points scored, can all be used for interesting applications, especially when combined with the course information. Several use cases will be presented in the Results section.

**Table 4**: Race Result Features. Abbreviated names between brackets for usage in Table 5.

| Feature | Description |
|---|---|
| **Race Name** | Name of the race. Can be matched exactly with GPX files by appending '.gpx'. |
| **Date** | Date in DD Month YYYY format (e.g., 03 April 2022). |
| **Rank** | End result rider in race. Numeric value or indicative of 'Did Not Finish' (DNF), 'Disqualified' (DSQ), 'Outside Time Limit' (OTL) or 'Did Not Start' (DNS). |
| **Team** | Team name of the team the rider represented. Note that these can change within one season due to transfers or national team representations. |
| **Name** | Rider name. |
| **Time** | Recorded time. First ranked rider always in absolute units. Relative time loss to first rider for all races besides team time trials (TTTs). TTTs have the exact time of the team reported. |
| **UCI points (UCI)** | UCI points scored through this single result. |
| **PCS points (PCS)** | PCS points scored through this single result. |
| **Team Time Trial (TTT)** | Indicator whether (1) or not (0) the race was a team time trial. |
| **TimeAfterTeamTTT (TAT)** | If not missing, this is the difference between the athlete and the main time of the team. |
| | Always missing in non-TTTs. |

**Table 5**: Example Output *race_results_2017_2023*: 2023 Craywinckelhof - Omloop van het Hageland. * Denotes '2023 Craywinckelhof - Omloop van het Hageland'. ** Denotes Team Name (i.e., Team SD Worx, UAE Team ADQ, etc.)

| Race Name | Date | Rank | Team | Name | Time | UCI | PCS | TTT | TAT |
|---|---|---|---|---|---|---|---|---|---|
| * | 26 Feb 2023 | 1 | ** | WIEBES Lorena | 3:21:19 | 125 | 75 | 0 | NA |
| * | 26 Feb 2023 | 2 | ** | BASTIANELLI Marta | 0:00 | 85 | 55 | 0 | NA |
| * | 26 Feb 2023 | 3 | ** | CORDON-RAGOT Audrey | 0:00 | 70 | 40 | 0 | NA |
| * | 26 Feb 2023 | 4 | ** | FIDANZA Arianna | 0:00 | 60 | 32 | 0 | NA |
| * | 26 Feb 2023 | 5 | ** | FOURNIER Roxane | 0:00 | 50 | 28 | 0 | NA |
| * | 26 Feb 2023 | 6 | ** | PATERNOSTER Letizia | 0:00 | 40 | 24 | 0 | NA |
| * | 26 Feb 2023 | 7 | ** | DE WILDE Julie | 0:00 | 35 | 20 | 0 | NA |
| * | 26 Feb 2023 | 8 | ** | BIANNIC Aude | 0:00 | 30 | 18 | 0 | NA |
| * | 26 Feb 2023 | 9 | ** | HANSON Lauretta | 0:00 | 25 | 16 | 0 | NA |
| * | 26 Feb 2023 | 10 | ** | DE ZOETE Mylène | 0:00 | 20 | 14 | 0 | NA |
| * | 26 Feb 2023 | 11 | ** | BORGHESI Letizia | 0:00 | 15 | 12 | 0 | NA |
| * | 26 Feb 2023 | 12 | ** | BOILARD Simone | 0:00 | 10 | 10 | 0 | NA |
| * | 26 Feb 2023 | 13 | ** | VAN ZUTHEM Cecilia | 0:00 | 5 | 8 | 0 | NA |
| * | 26 Feb 2023 | 14 | ** | GUTIÉRREZ Sheyla | 0:00 | 5 | 7 | 0 | NA |
| * | 26 Feb 2023 | 15 | ** | REUSSER Marlen | 0:00 | 5 | 6 | 0 | NA |
| * | 26 Feb 2023 | 16 | ** | JACKSON Alison | 0:00 | 3 | 5 | 0 | NA |
| * | 26 Feb 2023 | 17 | ** | VIGIE Margaux | 0:00 | 3 | 4 | 0 | NA |
| * | 26 Feb 2023 | 18 | ** | CHAPMAN Brodie | 0:00 | 3 | 3 | 0 | NA |
| * | 26 Feb 2023 | 19 | ** | VANPACHTENBEKE Margot | 0:00 | 3 | 2 | 0 | NA |
| * | 26 Feb 2023 | 20 | ** | VIGILIA Alessia | 0:00 | 3 | 1 | 0 | NA |
| * | 26 Feb 2023 | 21 | ** | MACKAIJ Floortje | 0:07 | 3 | NA | 0 | NA |
| * | 26 Feb 2023 | 22 | ** | TRUYEN Marthe | 0:07 | 3 | NA | 0 | NA |
| * | 26 Feb 2023 | 23 | ** | MARKUS Femke | 0:07 | 3 | NA | 0 | NA |
| * | 26 Feb 2023 | 24 | ** | DEMAY Coralie | 0:07 | 3 | NA | 0 | NA |
| * | 26 Feb 2023 | 25 | ** | UNEKEN Lonneke | 0:18 | 3 | NA | 0 | NA |

## 4. Results: Data Validation + Potential Applications

The GPX files can provide many interesting insights. Figure 4 (left pane) shows the geographical dispersion of the organized races, highlighting a clear geographic bias in top-level road cycling. Our data confirm Van Reeth's study[21] about the globalization trends in cycling, which concluded that the non-European representation in the calendar remains limited but constantly growing. Looking at the period 2017-2023, UCI cycling is mostly popular in Europe, with limited popularity in Northern South America (i.e., Colombia and Venezuela), Argentina, South-Eastern Asia, and the Middle-East. Identifying the key drivers behind the success of race organizers in these areas may be interesting avenues for feature research, and could help the sport in further establishing strategies related to its further globalization.

When focusing on Western Europe (Figure 4, right pane), we observe large regional differences: cycling is widely popular in Northern France and Belgium, with several interesting clusters in France, Italy, and Spain. In France, the sport is mainly popular in Bretagne and the French Alps. In Italy, it is mainly popular in Lombardy, Tuscany, and Veneto regions. Spanish race popularity is clearly dominated by the Basque region. Remarkably, neither Italy nor the Basque country have a top-level team (i.e., WorldTour) at this time, while there is a clear interest for the sport in these regions. This might be of relevance to potential firms interested in sponsoring with local commercial interests.

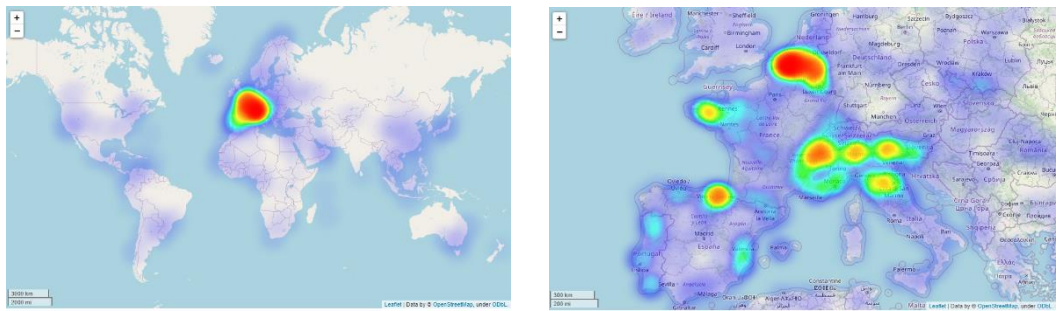

**Figure 4**: Heatmap of Global Geographical Locations (left pane) and Western European Geographical Locations (right pane).

Since GPX files may contain incorrect elevation data points, we use a digital elevation model[17] – the Shuttle Radar Topography Mission elevation data set – to check these values based upon the corresponding geographical location. Our results show that the elevation used in the GPX files is accurate, with only minor deviations from the digital elevation model. Overall, the average root mean squared error between original value and digital elevation model value is just 8.14 m. An in-depth inspection reveals that this is mostly due to some deviations in the digital elevation model, with major overlap between both values. Figure 5 (a + b) displays this for the 2017 Brussels Cycling Classic. The overlap between both data sources is remarkable, with the original GPX file data having the additional advantage to give more stable values. However, the issue becomes more serious when looking at cases where the maximal deviation between both sequences becomes quite large. Figure 5 (c + d) shows an example of this, with the 2017 Tour des Fjords Stage 3. Once again, we notice a more stable pattern in the left panel showing the original data, and a less stable sequence resulting from the digital elevation model. However, this time this is also worsened by a single extreme peak in the first half of the digital elevation model. Such a peak is also not present in the official course representation provided (see Figure 6). Similar cases could be found across our analysis, indicating a higher level of reliability of the original data compared to the data from the digital elevation model. Accordingly, we use the original values from the GPX files.

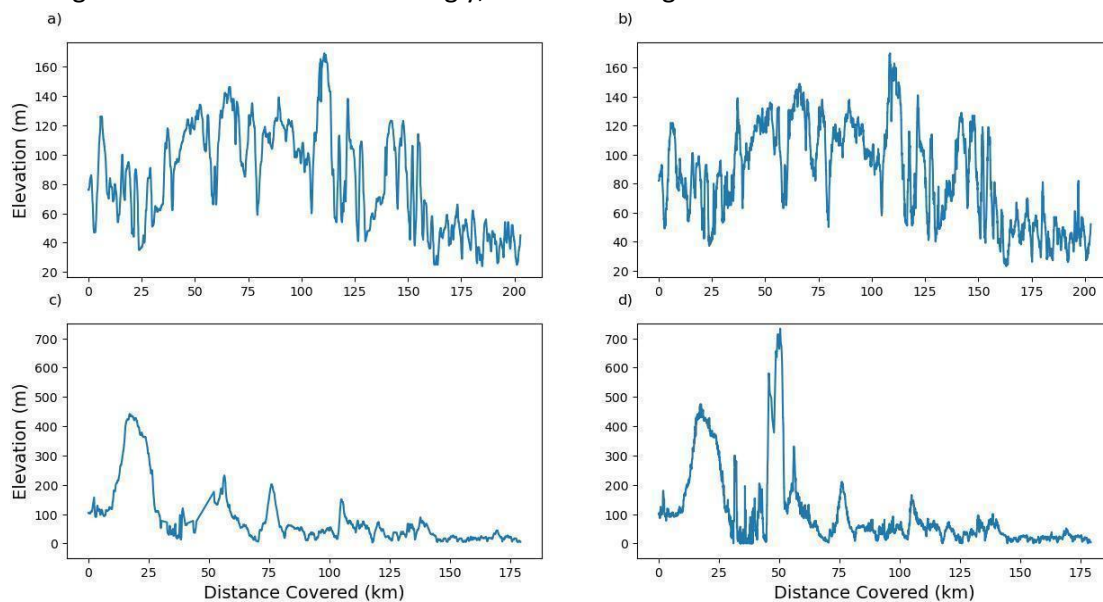

**Figure 5**: (a + b) Visualization Elevation Profile 2017 Brussels Cycling Classic with original data (a; left upper pane) and data from digital elevation model (b; right upper pane). Note the more shaky pattern of the digital elevation model. (c + d) Visualization Elevation Profile 2017 Tour des Fjords Stage 3 with original data (c; left lower pane) and data from digital elevation model (d; right lower pane). Besides the more shaky pattern of the digital elevation model, there is also a clear difference between both data sources between 25 and 50 km raced.

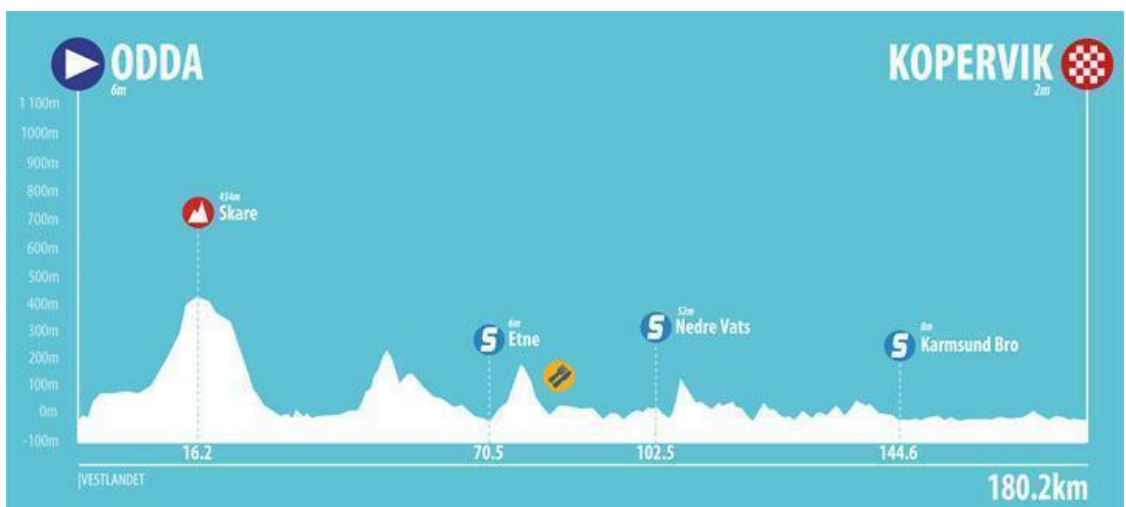

**Figure 6**: Officially communicated course profile of 2017 Tour des Fjords Stage 3. There clearly is no climb above 600m after the initial climb to around 500m elevation. Image from FirstCycling[22]

The information on the course profile allows us to learn which races are more related to others. Every cycling fan and expert will know that a race like the Ronde Van Vlaanderen will contain more information with regard to expected performance in the upcoming Paris-Roubaix race compared to how riders performed in the mountain stages of the Volta Catalunya. However, this has been completely disregarded by past research. Kholkine et al.[9] manually select relevant races per predicted race on a sample of six races. This is a heavy manual task which can hardly be translated to larger data sets, and will never cover the entire racing calendar. Our introduced data set lends itself perfectly for such a task: knowing the course characteristic of each race, it is now feasible to group races in a data-driven way. However, the strength of expert knowledge should also not be disregarded. Accordingly, we cluster the races based upon the features in the *structured_course_data* data set (i.e., all features besides the race name) using the constrained K-means clustering algorithm[23]. This way we can feed information on races which should definitely be clustered together (e.g., 2018 Ronde Van Vlaanderen and 2019 Dwars door Vlaanderen), and races which should never be linked together (e.g., 2018 Ronde Van Vlaanderen and 2021 Tirreno-Adriatico Stage 5) in the optimization process. This way we can both use our rich data set and background knowledge. A total of 149 pairs was annotated as '*must link*' or '*cannot link*', of which 49 are used as constraints, and 100 are used for a posterior external validation. Two imputation techniques will be compared: (1) zero imputation, where each missing value gets a value of zero, and (2) KNN imputation[24]. After imputation, standard scaling is applied.

The zero imputation would make a lot of sense from a theoretical perspective here as a missing value is not missing at random, but rather indicates that no value of this way or surface type was detected. This is also reflected in Figure 7, which depicts the performance of both imputation techniques at different settings of K on the hold out test set of 100 pairs (panel a), as well with regard to the internal validation (silhouette score; panel b). Overall, the performance is limited, with relatively low internal validation. Nonetheless, the external validation suggests a useful grouping. The silhouette score was used to determine the ideal K (i.e., K = 9 for zero imputation). The silhouette score was selected over the external validation, as the optimum showed results which were better reflections of expert judgement.

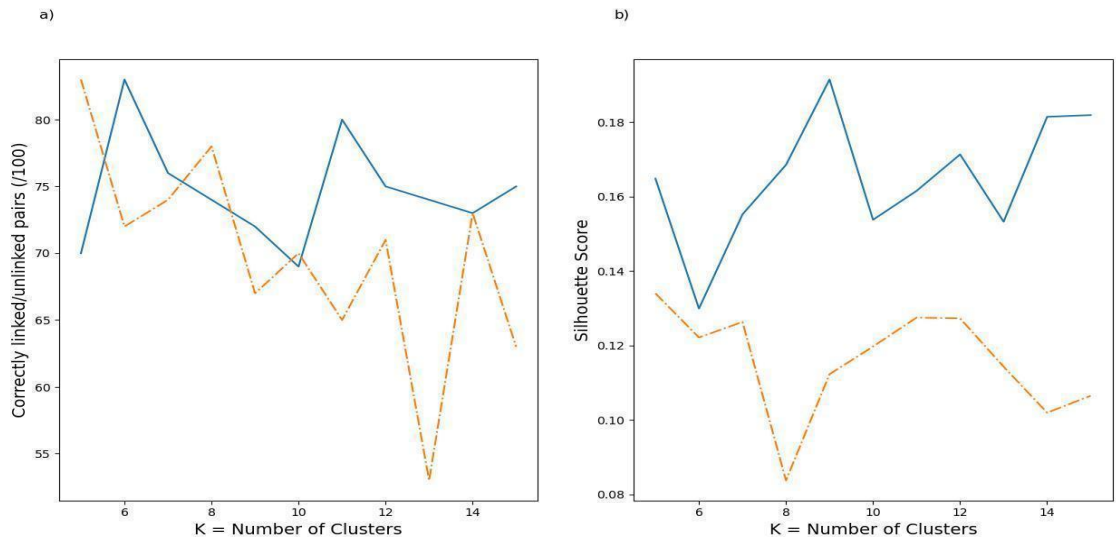

**Figure 7**: Comparison Imputation Techniques. Zero imputation (blue solid line) leads to better results compared to KNN imputation (orange dotted line), both with regard to external validation (test set validation; panel a), as well as to internal validation (silhouette score; panel b).

When inspecting the resulting 9 clusters (Table 6), we also observe several useful clusters to be formed. Some clear specializations are already present, which can be used for downstream tasks. However, further research can probably result in significant improvements, as the clusters seem to be focusing relatively much on surface and road type, while leaving out important specializations (e.g., steep hill top finishes, medium mountain stages). This indicates that future research on the data set may focus on feature engineering, determining which features are most relevant to determine race specialization. Interesting candidates could be the location (e.g., 10 km before finish) of hurdles such as climbs or non-asphalt sections. Moreover, notice the 'Missing Road Types' cluster. This cluster was mainly composed of races where all road and surface types had been imputed with zero due to the GPX not providing the correct information through our methodology. It is evident that these should not be truly clustered together, and it might be interesting to work out methodologies to handle this cluster separately. This is also nicely visualized in Figure 8, panel b, where you can observe the races from this cluster (brown dots) to have no relationship between distance and paved road, as the paved value is always missing, and imputed with zero. Notice how panel c visualizes that the cobble cluster (purple dots) is mainly driven by a significant cobble distance, and that panels d and e visualize how the mountain stages (orange dots) are mainly grouped together due to their higher vertical gain. It is also observed that the 'Missing Road Types' cluster has an undesirable heterogeneity.

**Table 6**: Identified Clusters

| Cluster | Number of Races |
|---|---|
| Sprint Stages | 2,070 |
| Short/Flat Races | 1,792 |
| Missing Road Types | 1,301 |
| Mountain Stages | 1,155 |
| Wide Roads | 1,067 |
| Cobble Races | 415 |
| High Elevation | 372 |
| Races Outside Europe | 220 |
| Races with Off-Road Sections | 97 |

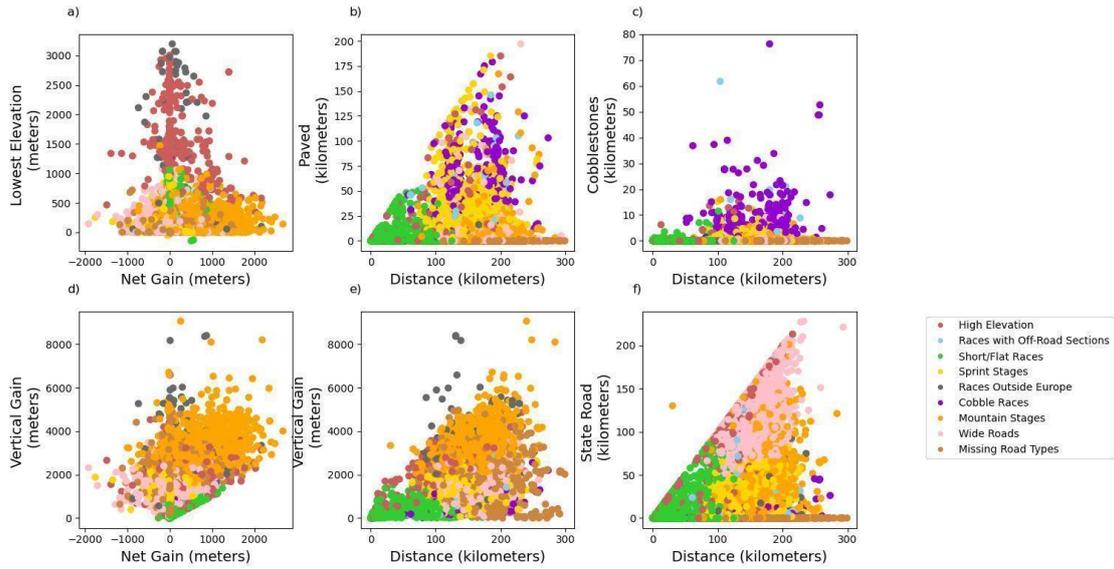

**Figure 8**: Visualization of Clusters along some important features.

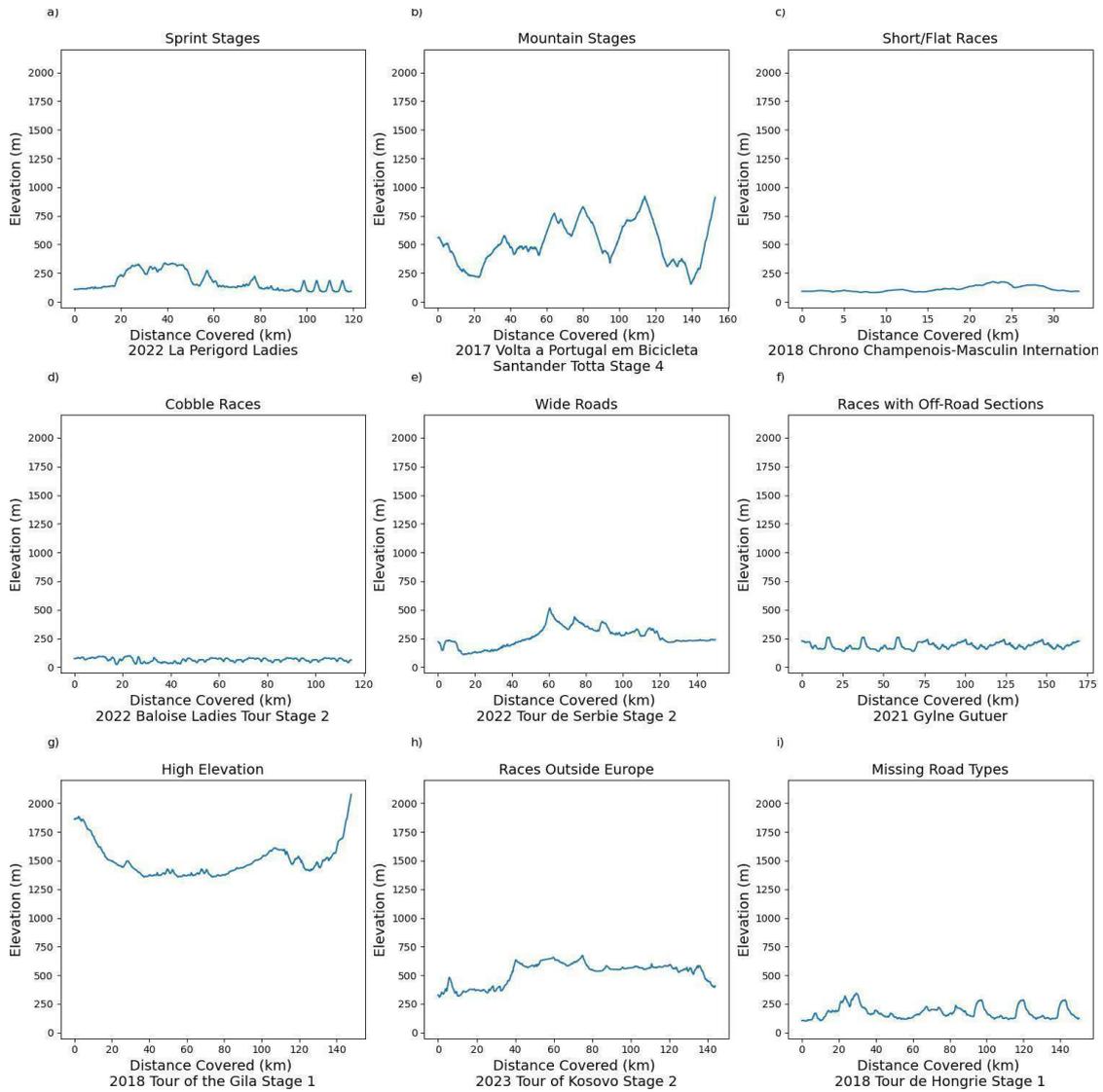

**Figure 9**: Visualization of Elevation Profile of Cluster Medoids. Clusters which are mainly driven by elevation and distance, such Mountain Stages (b), Short/Flat Races (c), and High Elevation (g) are most distinguishable from the Elevation Profiles.

To give some further insight into what drives each cluster, the medoid of each cluster is determined, and its elevation profile is depicted in the related panel of Figure 9. Logically, the clusters which are heavily determined by elevation and distance are most pronounced. Mountain Stages (8b) typically have several longer climbs, while the Short/Flat Races cluster (8c) typically contains time trials, much shorter than regular road races, and High Elevation (8g) distinguishes itself from the rest of the clusters by having very high Lowest Elevation values. Other clusters are harder to distinguish. The Sprint Stages cluster medoid (8a) actually has a relatively hilly final, which might indicate that this cluster (the largest one) is too broad, which may be alleviated by the above mentioned further focus on feature engineering in follow-up research. Interestingly, the Cobble Races medoid (8d) shows an elevation profile which would match the elevation profile of races happening in the Flemish Ardennes, a Belgian region known for hosting the majority of the cobbled classics. Other clusters show a less typical pattern, given the fact that their distinctiveness is more determined by road or surface type, which is not visualized by elevation profiles.

The results in **race_results_2017_2023** are of most value when combined with the linked data from the race courses. It is easy to link the previously determined clusters to the race results. An interesting first application could be to calculate the points scored per rider per cluster. The results of this small analysis are depicted in Table 7, which depicts the top-5 riders for three of our clusters in terms of PCS points. Interestingly, it shows both the potential of our data set, as well as potential for future improvements. The short/flat races cluster contains mostly time trials, resulting into the detection of all major male time trial starts. The mountain stage cluster is also coherent, leading to the top-5 riders all being world class climbers. Finally, the missing road type cluster has several all-rounders placed on top. This makes sense, given the more heterogeneous character of this cluster. Overall, the very simple methodology already enables us to uncover several interesting specialization evaluations due to the richness of our data set, which will probably be improved after more sophisticated clustering is performed. Future research can likely create multiple in-depth valuations of rider specializations.

**Table 7**: 5 Riders to score Highest Number of PCS Points in Various Clusters

| Cluster | Rider | PCS points |
|---|---|---|
| Short/Flat Races | GANNA Filippo | 1752 |
| Short/Flat Races | KÜNG Stefan | 1663 |
| Short/Flat Races | DENNIS Rohan | 1177 |
| Short/Flat Races | ROGLIČ Primož | 1167 |
| Short/Flat Races | EVENEPOEL Remco | 1130 |
| Mountain Stages | POGAČAR Tadej | 4646 |
| Mountain Stages | ROGLIČ Primož | 4189 |
| Mountain Stages | VALVERDE Alejandro | 2969 |
| Mountain Stages | ALAPHILIPPE Julian | 2383 |
| Mountain Stages | PINOT Thibaut | 2143 |
| Missing Road Types | KRISTOFF Alexander | 3784 |
| Missing Road Types | SAGAN Peter | 3681 |
| Missing Road Types | VAN AVERMAET Greg | 3671 |
| Missing Road Types | VAN AERT Wout | 3135 |
| Missing Road Types | VIVIANI Elia | 3021 |

The data set also holds large potential with regard to several race outcome prediction and team roster optimization methodologies. To demonstrate this, we built the following small case study. We selected Mark Cavendish, as he is a rider which affinities are easy to evaluate as a domain expert: he thrives during flat races, being one of the candidates for the end victory, while he is generally among the last to finish in the high mountains, often even fighting to finish before the time limit[25]. A predictive model (i.e., a random forest classifier) was built based on the races during which Cavendish participated in the period 2017-2022 to predict how likely he was to finish in the top-3 of the races he participated in during the 2023 season.

Independent features were the structured course features reported in Table 2. The very simple model achieved an AUC of 0.6192. Moreover, the model is able to pick up which races suit him best. Consider the races in Table 8: the model is clearly able to pick up the races which suit Cavendish best in the 2023 test season, with very flat or even overall downhill stages to be ideal (e.g., 2023 Giro d'Italia Stage 17; visualized in Figure 10a), while stages with heavy mountains or steep hills ideal (e.g., 2023 Tirreno-Adriatico Stage 5; visualized in Figure 10b) are indicated as being mall-fitted to his capabilities. For example, during the 2023 Giro d'Italia Stage 17, he only finished 19th, but the peloton was still very large, with Cavendish in it, and be super accurate in actual outcome prediction, it is able to learn how well the race would suit him.

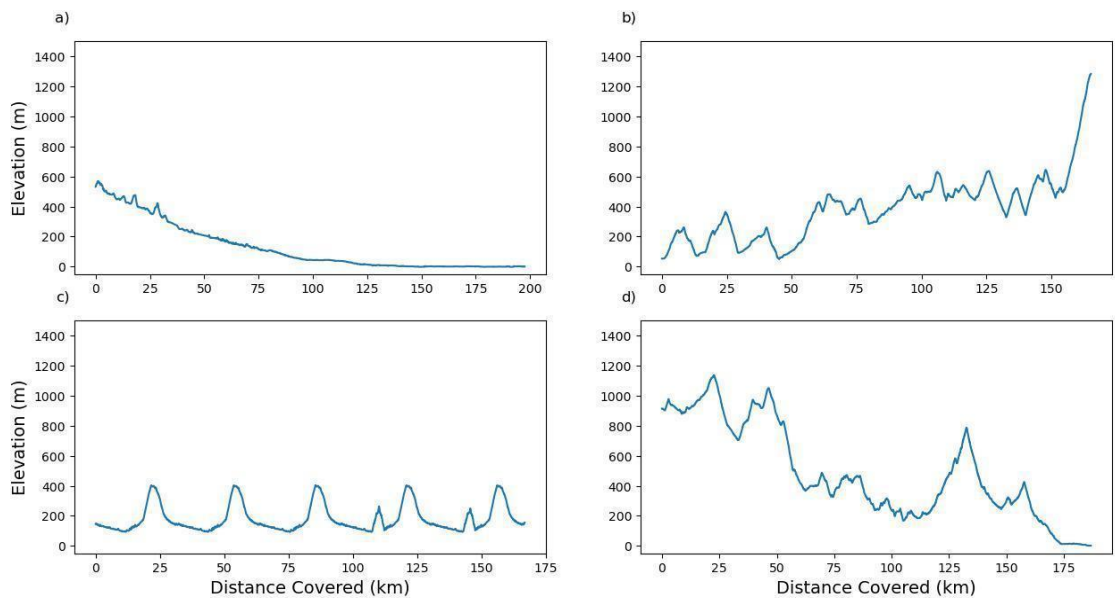

**Figure 10**: Visualization of Elevation Profile of Best- and Least-Suited Races for Mark Cavendish (a + b) and Juan Ayuso (c + d). Cavendish thrives on very easy profiles (2023 Giro dItalia Stage 17; a), and struggles in the high mountains (2023 Tirreno-Adriatico Stage 5; b). Ayuso's multi-capabilities show slightly less distinguished differences, but with a clear preference towards mountaintop finishes and the 2023 Prueba Villafranca - Ordiziako Klasika, a race which he has consistently thrived in in the past (c), and with the least preference going to races likely ending in a sprint due to the overall downwards slope (2023 La Vuelta Ciclista a Espana Stage 5; d).

Moreover, the methodology works across a wider range of riders. Consider the Spanish rider Juan Ayuso. At 21 years of age, he is much less experienced, and has a more limited list of results in our sample, with only 120 race results, of which 39 in the 2023 season. Learning his capabilities will inevitably be harder, as we only have 81 race results leading onto the 2023 season to predict his capabilities there. The young rider is less successful compared to Cavendish, and accordingly we have altered our definition of success as a top-10 placement. Test set performance is in line with our Cavendish model, with a test set AUC of 0.6467. Nonetheless, this simple approach can still learn which type of races he is more suited to, as reflected in Table 8, and visualized in Figure 10 (c + d). It is clear that Juan Ayuso already is conscious about the races he rides and is very versatile, with no stages having a very low probability of him ending up in the top-10. However, the model clearly gives much less probability to a generally downwards-sloping race like the 2023 La Vuelta Ciclista a Espana Stage 5 (Figure 10d), compared to races which feature more upwards slopes at the end of the race. Interestingly, four out of five races deemed most favorable to Juan Ayuso are mountain top finishes, while the race deemed most favorable (2023 Prueba Villafranca - Ordiziako Klasika; Figure 10c) is a hilly circuit race, a race in which Ayuso placed 2nd and 4th during previous participations. The course of this race is the same across years. It seems that the model overfits to these exact observations during the training period, learning that the race suits him very well, and that the general trend favors mountain top finishes.

Once again, the course affinity of the rider can be determined with a simple analysis, using a limited sample of observations due to the linking between geographical course profile information and race results as we present with our data set. This can be used for teams to screen up front how well each rider would be suited to each individual race, and use this to optimize their race schedule, as well as their rider-race assignments.

**Table 8:** Race Affinities Mark Cavendish + Juan Ayuso in 2023 test season

| 5 Most Favorable Races | Podium Probability | 5 Least Favorable Races | Podium Probability |
|---|---|---|---|
| MARK CAVENDISH | | | |
| 2023 Giro dItalia Stage 17 | 0.35 | 2023 Tour of Oman Stage 5 | 0.01 |
| 2023 Tour of Oman Stage 1 | 0.33 | 2023 Giro dItalia Stage 12 | 0.01 |
| 2023 ZLM Tour Stage 3 | 0.30 | 2023 Tour de France Stage 5 | 0.01 |
| 2023 Presidential Cycling Tour of Turkiye Stage 1 | 0.30 | 2023 E3 Saxo Classic | 0.01 |
| 2023 Giro dItalia Stage 20 | 0.26 | 2023 Tirreno-Adriatico Stage 5 | 0.01 |
| JUAN AYUSO | | | |
| 2023 Prueba Villafranca - Ordiziako Klasika | 0.83 | 2023 La Vuelta Ciclista a Espana Stage 4 | 0.31 |
| 2023 La Vuelta Ciclista a Espana Stage 3 | 0.81 | 2023 La Vuelta Ciclista a Espana Stage 19 | 0.30 |
| 2023 Tour de Romandie Stage 4 | 0.75 | 2023 La Vuelta Ciclista a Espana Stage 7 | 0.24 |
| 2023 La Vuelta Ciclista a Espana Stage 6 | 0.74 | 2023 Donostia San Sebastian Klasikoa | 0.19 |
| 2023 La Vuelta Ciclista a Espana Stage 13 | 0.71 | 2023 La Vuelta Ciclista a Espana Stage 5 | 0.19 |

## 5. Discussion

## 5.1. Standardization

A first limitation of current literature which our open data set resolves is the lack of standardization. Each prior study in the field uses a unique data set which makes it very hard to compare the various proposed methods. For example, [10] and [11] both create machine learning-based automated scouting systems. However, it is infeasible to compare both works as the data sets used are not publicly available. This has hampered the development of the field of cycling analytics. In another study [26], a new methodology is deployed to measure individualized rider performance across many races and participants lists. In the next phase of this research, this measurement is used as a feature engineering step for race outcome estimation, and benchmarked against the methodology in [9]. However, once again this needs to be deployed on other races, and the entire methodology from [9] needs to be replicated. Sharing the data and code in these studies would lead to an increased level of replicability, which could attract researchers from related fields. Related fields, such as soccer analytics, have seen similar boosts in popularity after the publication of open and easy-to-access data sets such as [27].

## 5.2. Geographical Course Information

The major contribution of this data set should be situated in the linking of the race results with the race profile information through the GPX files. This linking can be automated but needs a lot of tedious manual checking of the web scraper to ensure no false linkages. This might explain why this precious source of information is currently disregarded in the field. Throughout Section 4, we give several examples of how this data can lead to various new insights, which also bears the potential for end applications when developed further. For example, when comparing Mark Cavendish and Juan Ayuso (Table 8 and Figure 10), we observe very different races to be suggested as most suited to the two respective riders. This observation will come as no surprise to stakeholders in the industry, as these specializations are widely acknowledged. However, there is almost no accountancy for this throughout cycling analytics literature, and no accountancy in an automated way. This data set provides modelers the option to account for this without having to go through the cumbersome process of collecting this data manually. We are confident that this will boost the number of researchers working in the field, and we are curious to see the developments made in the near future.

## 6. Conclusion

This paper and corresponding analyses show that the data set is very rich and allows for advanced analyses uncovering the key role geography plays in road cycling and road cycling performance. Several initial avenues for research are explored. We demonstrate how rudimental feature engineering already allows for a useful data-driven grouping of races, a practice which currently is still being done manually[9,11]. Future research can delve deeper in feature engineering, or can explore several methodologies which automate this process like recurrent neural networks, to which the sequential nature of the race courses seems well-suited.

It is also shown that the data allows to learn the relationship between athlete, course profile, and performance. In our preliminary methodology, we use an athlete-by-athlete approach. Future solutions might focus on learning this relationship across athletes, using rider representations, which would allow to learn more complicated relationships, and a deeper understanding of this complex relationship. This would allow for a fully data-driven team roster optimization. These analyses are just the tip of the potential analytics iceberg. The data set allows for very rich geospatial cycling analytics. One can imagine several interesting end-applications. For instance, one could learn the relationship between race course and race outcome (e.g., how many people finished in the top group). When combined with literature on alternative routing[28], this could allow race organizers to come up with race courses much quicker.

Finally, we do not claim that this data set contains all feasible data sources possible in professional road cycling. However, other sources are difficult to obtain in an exhaustive way due to privacy regulations (e.g., athlete sensor data) or due to issues with ownership (e.g., video broadcasts). Accordingly, we believe that our data set is a useful addition to the literature on cycling analytics, and that it allows for standardized benchmarking in the field, which was currently infeasible. Moreover, the data set allows users to incorporate fine-grained information on the race course. Currently, this information was almost completely disregarded or used on a very coarse level of information. Our data set allows researchers to also use this data on a very detailed level without the threshold of having to collect and link these sources which, as discussed throughout this manuscript, is a cumbersome process.

## Code Availability

The code to reproduce the plots in the paper is publicly available at https://github.com/bram-janssens/CyclingAnalytics.

## Acknowledgements

This work has been partially supported by EU project H2020 SoBigData++ G.A. 871042 through the Transnational Access program. Additional funding has been granted through Ghent University's Special Research Fund (BOF) [BOF/STA/202009/001] and the Research Foundation Flanders' (FWO) postdoctoral fellowship [12ZM923N].

## Author contributions

Dr. Bram Janssens: Data collection, data analysis, conceptualization, writing first draft; Dr. Luca Pappalardo: Conceptualization, feedback; Dr. Jelle De Bock: Data collection; Prof. Dr. Matthias Bogaert: Feedback; Prof. Dr. Steven Verstockt: Feedback.

## Competing interests

The authors declare there is no conflict of interest.